\documentclass[aip,reprint]{revtex4-2}

\usepackage{amssymb}
\usepackage{graphicx}
\usepackage{dcolumn}
\usepackage{bm}
\usepackage{longtable}
\usepackage{color}
\usepackage{hyperref}

\newcommand{\cm}{cm$^{-1}$}

\newcommand{\Sch} {Schr\"{o}dinger}

\begin{document}

\preprint{} 

\date{\today}

\title{Description of the condensed phases of water in terms of Bose-Einstein condensates}

\author{Fran\c{c}ois Fillaux}

\email[]{francois.fillaux@sorbonne-universite.fr}
\affiliation{Sorbonne Universit\'{e}, CNRS, MONARIS, 4 place Jussieu, Paris, F-75005 France.}


\begin{abstract}
I show that the reason why many properties of water are notoriously at odds with current models is the low dissociation energy of the H-bond, which leads to the instability of the Maxwell-Boltzmann distribution. Bose-Einstein condensation at the gas-liquid transition stabilizes the H-bond and opens the door to a reasoning from the laws of quantum mechanics to classical physics, explaining the properties of the condensed phases with only four observables and no arbitrary hypotheses. The isomorphic structures are composed of honeycomb sheets. Heat transfer reveals the eigenstates that determine the temperatures of the phase transitions. Quantum entanglement and degeneracy explain the heat capacities and latent heats. The supercooled liquid is a superposition of the high and low density eigenstates. Bose-Einstein condensation also explains the honeycomb pattern of an aerosol of droplets. 
\end{abstract}

\maketitle

\begin{figure}
\begin{center}\includegraphics[angle=0.,scale=0.5]{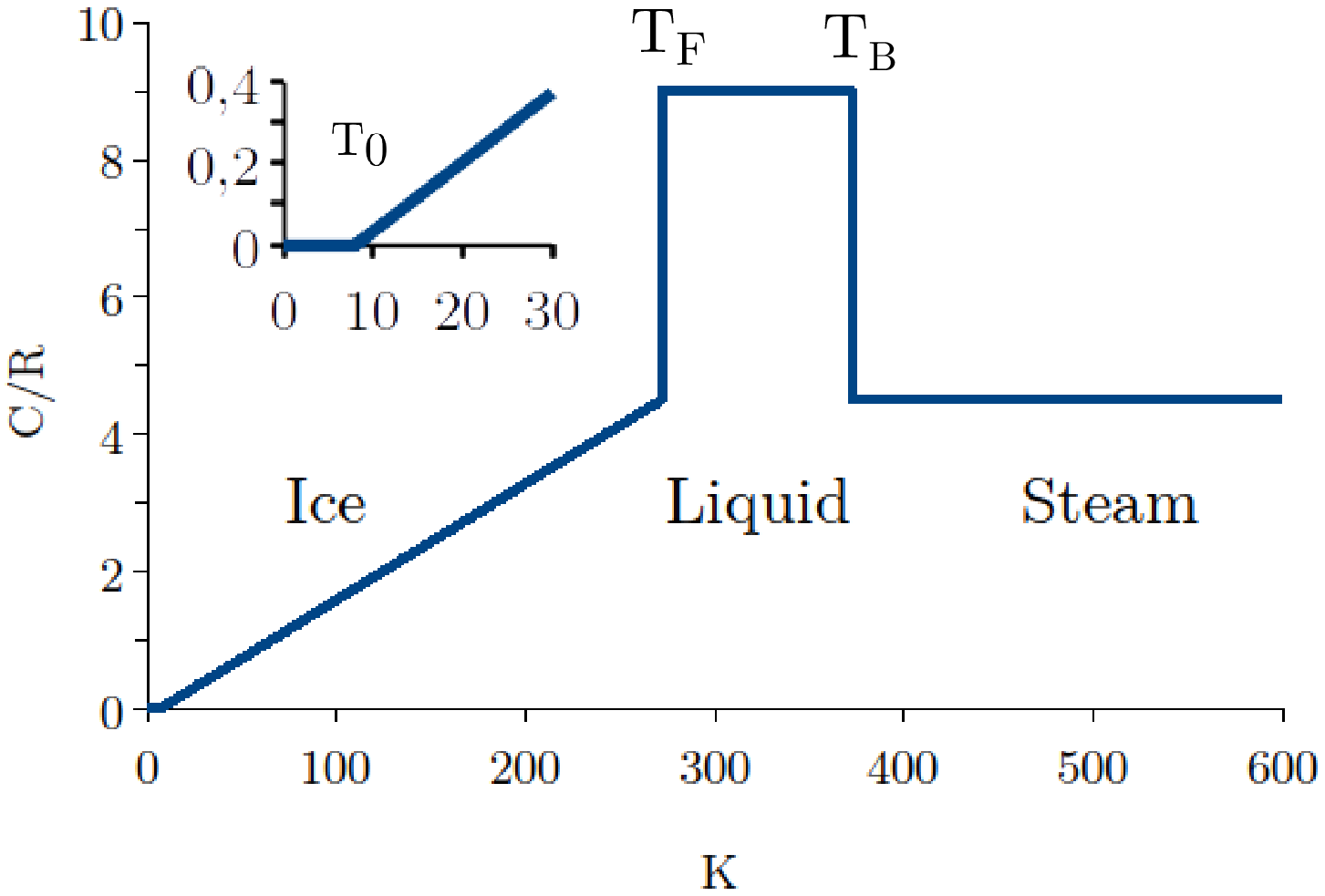}
\caption{\label{fig:1} Reduced molar heat-capacity $C/\mathcal{R}$ of H$_2$O (Table \ref{tab:1}). }
\end{center}\end{figure}

Water, the matrix of life, covers two-thirds of our planet and is one of the most abundant substances in the universe. It is of paramount importance in physics and chemistry, earth and life sciences, cosmology and technology. No other molecule exists as a solid, liquid, or gas at normal pressure. At the microscopic level, the most popular descriptions are based on Maxwell-Boltzmann statistics for H$_2$O molecules subject to local forces. \cite{Ball,Chaplin2,Ice} In simple terms, ice Ih is a frustrated hexagonal lattice containing an exponential number of proton configurations according to the ``ice rules,'' \cite{BF,Pauling2,BSS2} liquid water is a tetrahedral network of cooperative H-bonds in a jumble of molecular clusters constantly breaking and forming, and vapor is composed of dimers linked by transient H-bonds. However, these descriptions do not capture nuclear quantum effects. \cite{WCD,CDK,KS,BKPS,PSA,SRA,Fil8}

Although water is probably the most studied substance, there is still no satisfactory theory based on an argument from the quantum laws of microscopic physics to macroscopic phenomena that is convincing in all respects, without arbitrary assumptions. The lack of theory brings modeling to the fore, but models are hardly conclusive, for they are notoriously sensitive to how forces are defined, and the large number of competing models indirectly underscores their lack of success. \cite{SBC,Ball,GY} The physics underlying water phases is still elusive, and this is a serious drawback for capturing the functioning of biological systems. \cite{Ball2} 

This paper presents a quantum description of the phases of water that explains thermal properties called ``anomalous'' because they are quite different from those found in other materials. This is the case for the heat capacity, which challenges two pillars of matter science: the Maxwell-Boltzmann statistics and the energy equipartition. In Fig. \ref{fig:1} and Table \ref{tab:1}, it can be seen that $C_W \approx9\mathcal{R}$ is independent of $T$ for the liquid, in violation of statistical laws, and is halved at the freezing point $T_F$ and the boiling point $T_B$, in violation of the equipartition theorem. For ice, $C_I = 0$ below $T_0$ does not fit Debye's $T^{3}$ law. \cite{SLL} For steam, $C_S = \frac{9}{2} \mathcal{R}$ is inconsistent with both statistics and equipartition. Another anomaly (not shown) is the dramatic increase in the heat capacity of supercooled water down to the temperature of homogeneous crystallization $T_H \approx226$ K. \cite{Ice,AOS} 

Below I show that quantum correlations can freeze the kinetic energy and halve the heat capacity (see Sec. \ref{sec:1}), and Bose-Einstein condensation (BEC) can cancel quantum and statistical fluctuations (see Sec. \ref{sec:2}). 

Bose-Einstein condensation of monoatomic systems is a critical phenomenon at cryogenic temperatures, but Fig. \ref{fig:1} shows that the condensation of water molecules is not. The condensation can be explained by the low dissociation energy of the H-bonds, which is $D_0 = (1105 \pm 10)$ \cm\ measured for the dimer $\mathrm{(H_2O)_2}$ in a molecular jet. \cite{RCM} (The calculated potential energy surface shows that $D_0$ essentially corresponds to doubly H-bonded dimers. \cite{SWK}) Knowing that the dimer dynamics are described by proton modes above $1600$ \cm, librations in the range 400-700 \cm, and $\mathrm{O\cdots O}$ translations below 200 \cm, \cite{Ice} it 
follows that the dissociative proton modes and the predissociative librations are not stationary. (This is evidenced by extremely broad infrared absorption bands. \cite{Ice}) Therefore, at the gas-liquid transition the dimers cluster in a stationary state below $200$ \cm, where the $\mathrm{H_2O}$ behave as rigid bodies by adiabatic separation. The Bose-Einstein condensation can explain the ``abnormal'' properties of water, because the thermal properties are markedly different from those prescribed by the laws of thermodynamics: the absolute temperature $T$ is not an internal variable and the entropy related laws do not apply (see Sec. \ref{sec:2}). 

\begin{center}
\begin{table}  
\caption{\label{tab:1} Reduced molar heat capacities of the phases of water. $\mathcal{R} \approx 8.314$ J.mol$^{-1}$.K$^{-1}$. $T_0 = (8 \pm 1)$ K. $T_F \approx 273.16$ K. $T_B \approx 373.16$ K. *Neutron scattering demonstrates ice Ih at 5 K. \cite{BKPS,PSA,SRA}}
\begin{tabular}{llllll}
& & $2C/(9\mathcal{R})$ & Ref. \\ 
\hline\\
Vapor & $T_B \leq T$ & $1.001$ & \onlinecite{Verma} \\
Liquid & $T_F \leq T \leq T_B$ & $2.02$ & \onlinecite{LMM} \\
Ice Ih & $T_0 \leq T \leq T_F$ & $1.01 \displaystyle{\frac{T-T_0}{T_F-T_0}}$ & \onlinecite{MK,FW} \\
Ice Ih* & $0 \leq T \leq T_0$ & $< 10^{-2}$ & \onlinecite{SLL} \\
\end{tabular}
\end{table}
\end{center}

\section{\label{sec:1} Dimers}

Since there is no fundamental theory for H-bonded dimers $\mathrm{HO_d-H \dots O_aH_2}$ composed of a donor $\mathrm{H_2O_d}$ and an acceptor $\mathrm{H_2O_a}$, the description relies on models of either classical or quantum nature.

The classical model consists of a dimensionless proton in an asymmetric double well along the O$\cdots$O coordinate. The asymmetry is the energy difference between the $\mathrm{HO_d-H \dots O_aH_2}$ (L) and $\mathrm{HO_d \dots H-O_aH_2}$ (R) configurations with opposite electric dipole moment ($\mu$) orientations, so that proton transfer and $\mu$ flip occur simultaneously. Using this model, Bove \textit{et al.} \cite{BKPS} fitted the inelastic neutron scattering (INS) spectra of ice at different $T$ with quasi-elastic profiles and derived relaxation rates supposedly due to thermally activated over-barrier jumps. However, they found that the absence of a temperature effect was inconsistent with the model and ultimately concluded that quantum effects were likely, but they did not pursue this lead. 

In quantum mechanics, a dimer in isolation is a superposition of $L$ and $R$ eigenstates that describe a double H-bonded dimer $\mathrm{HO \underline{\cdots} HH \underline{\cdots} OH}$ for which $\langle\mu_L + \mu_R \rangle =0$. \cite{Anderson} The translational states are stationary and the water phases correspond to three different types of superposition. 

Type \textbf{I}: In ordinary high density liquid (HDL, $T_F \leq T \leq T_B$) there is no dipolar interaction. The translational states can be represented as \cite{FTP,FCou5} 
\begin{equation}\label{eq:1} \begin{array}{rcl}
|\psi_{0}\rangle & = & \frac{1}{\sqrt{2}}[\cos \phi |\psi_{L}\rangle + \sin \phi |\psi_{R}\rangle\\
& + & i(\cos \phi |\psi_{R}\rangle + \sin \phi |\psi_{L}\rangle)];\ E_{0};\\
|\psi_{1}\rangle & = & \frac{1}{\sqrt{2}}[\sin \phi |\psi_{L}\rangle - \cos \phi |\psi_{R}\rangle\\
& + & i(\sin \phi |\psi_{R}\rangle - \cos \phi |\psi_{L}\rangle)];\ E_{0}+\hbar\omega_{1}.\\
\end{array}\end{equation}
$|\psi_{L} \rangle$ and $|\psi_{R} \rangle$ are the zero order localized states, and $\hbar \omega_1 = |E_{L0} - E_{R0}|$ is the asymmetry gap. $\phi \ll \pi$ determines the relaxation rate. This combination cancels out displacements perpendicular to $\mathrm{O\cdots O}$ and restricts the dynamics to 1-D. A drawback of this approximation is that it does not account 
for the shift of the maximum probability density to longer distances in the upper state. 

Type \textbf{II}: In supercooled low density liquid (LDL), $i \rightarrow 1$ in (\ref{eq:1}). Antiparallel dipoles in the ground state reduce the energy to $E_{0} - \hbar\omega_\mu$. The O$\cdots$O bond length is stretched by the larger asymmetry gap and the density is reduced. The upper state with parallel dipoles is unstable, so the upper stationary state is at $E_{0} +\hbar\omega_{1}$. 

Type \textbf{III}: In ice $i \rightarrow \pm 1$ in (\ref{eq:1}). Entanglement yields:
\begin{equation}\label{eq:2} \begin{array}{rcl}
|\psi_{0\pm}\rangle & ; & E_0 + \hbar(-\omega_\mu \pm \frac{1}{2} \omega_t); \\
|\psi_{1\pm}\rangle & ; & E_0 + \hbar(\omega_1 \pm \frac{1}{2} \omega_t); \\
\end{array}\end{equation}
where the $\psi_{\pm}$ are symmetric and antisymmetric combinations of the zero order wavefunctions, and $\omega_t \approx 2\phi \omega_1 \ll \omega_\mu$ is the tunneling splitting. Quantum beating describes stationary oscillations of both the dipolar orientation and the proton probability density. In the momentum representation, the Fourier transforms $FT\psi_{0+} = FT\psi_{1+}$ and $FT\psi_{0-} = FT\psi_{1-}$. The kinetic energy is $T$ invariant and equipartition is excluded. 

\textbf{I} and \textbf{III} explain neutron scattering data. 

First, neutron Compton scattering (NCS) probes the mean kinetic energy of the protons and the temperature law expected for equipartition is $\bar{E}(T) = \bar{E}_0 + \frac{3}{2} k_B T$, where the zero point energy is practically $T$ independent, $k_B$ is the Boltzmann constant, and $\frac{3}{2} k_B \approx0.12$ meV.mol$^{-1}$.K$^{-1}$. However, the observed law is quite different. \cite{SRA} For $T \leq T_F$, $\bar{E} = (153 \pm 2)$ meV.mol$^{-1}$ is practically a constant, because the kinetic energy of \textbf{III} is frozen in ice. For $T_F \leq T \leq T_B$, $\bar{E}(T) \approx \bar{E}_0 + \frac{3}{2}k_B (T-T_F)$ results from the equipartition of \textbf{I} in the liquid and the freezing of the kinetic energy at $T_F$. 

Second, for INS the question is whether the spectra consist of the broad quasi-elastic profile considered by Bove \textit{et al.} \cite{BKPS} or, alternatively, tunneling transitions partially overlapped by the elastic peak. Prima facie, the spectra may appear ambiguous. However, the absence of a temperature effect, \cite{BKPS} the split probability density of the protons, \cite{KLe} the NCS data, \cite{SRA} the heat capacity, and the quantum nature of the protons definitively rule out the classical model. Therefore, the intensity humps observed at $\pm (0.10 \pm 0.01)$ meV can be attributed without conflict to tunneling splitting $\hbar\omega_t/{k_B} = (1.2 \pm 0.2)\ \mathrm{K}$, with semi-subjective error bars. 

\begin{table} 
\caption{\label{tab:2} Energies, $\mathcal{E}$, and partition coefficients, $\Theta$, of the phases of water. $T_0 \approx 8$ K; $T_F \approx 273$ K; $T_B \approx 373$ K; $T_{H} \approx 226$ K; $T_{MD} \approx 277$ K. $\Omega_{HD} = 14(\frac{3}{2})^2$. SC: supercooled. }
\begin{tabular}{llll}
& & $\mathcal{E/R}$ & $\Theta$ \\
\hline 
\\
$T_B \leq T \leq T_c$ & Steam & $ \displaystyle{\frac{9}{2}} (T_B + T)$ & ---\\
\\
$T_{MD} \leq T \leq T_B$ & Liquid & $9 T$ & $\displaystyle{\Theta_{HD} = \frac{T-T_F}{T_B-T_F}}$\\
\\
$T_{H} \leq T \leq T_{MD}$ & SC & $(9 + \Theta_{SC}^2 \ln {\Omega_{HD}} ]T$ & $\Theta_{SC} = \displaystyle{\frac{T_{MD} -T}{T_{MD} - T_H}}$ \\
\\
$T_0 \leq T \leq T_F$ & Ice Ih & $\displaystyle{\frac{9}{2}} (T_F + T) $ & $\Theta_I = \displaystyle{\frac{T-T_0}{T_F - T_0}}$ \\ 
\\
\end{tabular}
\end{table} 

\begin{table} 
\caption{\label{tab:3} Critical temperatures and microscopic observables. $\hbar\omega_{t} / k_B = (1.2 \pm 0.2)$ K; $\hbar\omega_{1} / k_B \approx 129$ K; $\hbar\omega_\mu / k_B \approx 47$ K. The fraction $1/7$ is the 7-fold quasiboson footprint. }
\begin{center}
\begin{tabular}{llll}
$\hbar\omega _t / k_B$ & $\hbar\omega_1 / k_B$ & $\hbar(\omega_1 + \omega_\mu)/ k_B$ & $\hbar \omega_\mu/ k_B$\\
\hline \\
$\displaystyle{\frac{T_0}{7}}$ & $\displaystyle{\frac{9}{7}(T_B-T_F)}$ & $\displaystyle{\frac{9}{14} T_F}$ & $T_F - T_H$ \\
\end{tabular}
\end{center}
\end{table} 

\section{\label{sec:2} The phases of Water}

Consider a macroscopic number, $N$, of molecules at normal pressure in a box in diathermal equilibrium with a black body at $T$. The density is phase dependent. Boundary effects are negligible. The singlet and triplet spin states are degenerate. 
The spatial coordinates of the O-atoms are $\mathbf{r}_i$ ($i =1 , 2 \cdots N$). 

In the condensed phases, the wavefunction $\Phi (\mathbf{r}_1 \mathbf{r}_2 \cdots \mathbf{r}_{N} \mathrm{, } t)$ 
is symmetric with respect to the exchange $\mathbf{r}_i \rightleftarrows \mathbf{r}_j$ of any two coordinates, so nearest neighbors are equidistant in a hexagonal lattice composed of honeycomb layers. The unit cell consists of 7 indistinguishable molecules equally involved in 7 double H bonds $\mathrm{\underline{\cdots} O\underline{\cdots} HH\underline{\cdots} O \underline{\cdots}}$, and the wavefunction is symmetric with respect to the exchange of any two pairs of correlated H. The notion of a local H-bond must be abandoned, because each node is part of 6 hexagons, so each H can be assigned to 6/4 O-orbitals, and thus the degeneracy is $(\frac{3}{2})^2$. 

The dynamics is described with the order parameter $\Psi (\mathbf{r}, t) = \sum_{K(D_0)} \sqrt{N_{k} } \psi_{k}(\mathbf{r}) \exp i\omega_{k} t$, where $k \in K(D_0)$ and $K$ is the number of thermally accessible stationary states. $\psi_{k}(\mathbf{r}) \exp i\omega_{k} t$ is the dimer wavefunction, $N_{k}$ is the occupancy, $\sum_{K(D_0)} N_k = N$, and $\mathcal{E} = \sum_{K(D_0)} N_k \hbar\omega_k$. \cite{PO,Leggett2} Apart from its normalization, $\Psi (\mathbf{r}, t)$ is the \Sch\ wavefunction, but unlike the single dimer, which is subject to thermal and quantum fluctuations, $\Psi (\mathbf{r}, t)$ characterizes the behavior of a macroscopic number of dimers and can be considered a classical quantity. \cite{Leggett2} 

The internal energies are presented in Table \ref{tab:2}, and Table \ref{tab:3} gives the relations between the critical temperatures and the microscopic observables. 

\subsection{\label{sec:2.1} Ice Ih}

The empirical relation $k_B T_0 \approx7 \hbar\omega_t $ (Tables \ref{tab:1} and \ref{tab:3}) gives the tunneling gap of the honeycomb cell composed of 7 inseparable dipoles \textbf{III} probed by thermal waves, while INS probes single protons. Compared to (\ref{eq:2}), the eigenenergies are multiplied by 7 and the eigenfunctions are unchanged. 

The ground state is independent of time and $\mu \equiv 0$ means that thermalization is impossible. This state is not occupied and $\mathcal{R} T_0$ is the equilibrium state when the thermal contact is off. 

The order parameter for $T_0 \leq T \leq T_F$ is: 
\begin{equation}\begin{array}{l}\label{eq:3} 
\Psi_I (t) = \sqrt{N_{0-}}\psi_{0-} e^{i\omega_{tI}t} + \sqrt{N_{1+}}\psi_{1+} e^{i\omega_{1+I}t} \\
+ \sqrt{N_{1-}}\psi_{1-} e^{i\omega_{1-I}t} ;\\ 
\displaystyle{\frac{N_{0-}}{N_{7}}} = \displaystyle{ 1-\Theta_I} ; \\
\displaystyle{\frac{N_{1+} + N_{1-}}{N_{7}}} = \displaystyle{\Theta_I}; \\
\end{array}\end{equation}
$N_{7} = N/7$; $\omega_{tI} = 7\omega_t$; $\omega_{1+I} = 7(\omega_1 + \omega_\mu)$; $\omega_{1-I} = 7(\omega_1 + \omega_\mu +\omega_t)$. $\Theta_I$ is the occupancy partition coefficient (Table \ref{tab:2}). $|\Psi_I (t)|^2$ describes the quantum beats at $\omega_{BI1} = \omega_{1-I} - \omega_{1+I}$ and $\omega_{BI2} = \omega_{1+I} -\omega_{tI}$. The heat transfer by photons $\hbar \omega_{BI1}$ gives the heat capacity $C_I = \frac{9}{2}\mathcal{R} \Theta_I$ proportional to ($T-T_0$) (Table \ref{tab:2}). According to Plank's law, the relative power radiated at $\omega_{BI2}$ is $(\omega_{BI2}/\omega_{BI1})^3 \sim 10^{-9}$. The contribution of photons $\hbar \omega_{BI2}$ to the heat capacity is insignificant, but it is the key to thermalization. 

The degeneracy $\Omega_I = (\frac{3}{2})^2$ explains the diffuse background in the neutron diffraction pattern. \cite{NK} 

\subsection{\label{sec:2.2} Liquid water}

Fusion occurs at $T_F$ when the non-stationary states become thermally accessible. Thermal waves disentangle 7 inseparable pairs of dipoles \textbf{III} into 7 separable pairs of dipoles \textbf{I}. The HDL degeneracy is $\Omega_{HD} = 14(\frac{3}{2})^2$ and the heat of fusion is: 
\begin{equation}\label{eq:4}
\Delta H_F = \mathcal{R}T_F \ln \frac{\Omega_{HD}}{\Omega_{I}} \approx5993\ \mathrm{J.mol}^{-1}.
\end{equation}
This compares favorably with the empirical value of $\approx6005\ \mathrm{J.mol}^{-1}$. \cite{Ice,FW} There is no significant contribution from dissociation, so the number of H-bonds per unit is unchanged. If the $\approx0.2\%$ overshoot of the empirical value is significant, a simple explanation is that crystallization can be delayed to $T_F - \Delta T$ by the absence of fluctuations. The decrease of the frozen kinetic energy, $\frac{9}{2} \mathcal{R} (T_F - \Delta T)$, causes an increase in the heat of fusion at $T_F$, so the empirical value taken at face value gives $\Delta T \approx0.3$ K. 

The HDL eigenenergies are $\mathcal{R} T_F$ and $\mathcal{R}T_B$, and the partition coefficient is $\Theta_{HD}$ (Tables \ref{tab:2} and \ref{tab:3}). 

Among the $12(\frac{3}{2})^2$ degenerate H-bonds per hexagon, a fraction $X_{LD} = [12(\frac{3}{2})^2 - 1]^{-1} \approx0. 038$ has identical orbital configurations and antiparallel dipoles. These are the precursors of the non-degenerate dimers \textbf{II}, which are the constituents of LDL. So the LDL eigenenergies are $\mathcal{R} T_{H} = \mathcal{R} (T_F - \hbar\omega_\mu/k_B)$ and $\mathcal{R} T_B$ (Table \ref{tab:3}). For $\mathcal{R} T_{H}$, $\mu = 0$ means that this state is insulated from thermal waves. 

By cooling HDL, $\mathcal{R}T_H$ remains unoccupied as long as the $\mathcal{R}T_B$ occupancy is dominant. The bifurcation occurs at $T_{MD}$, where the $\mathcal{R}T_B$ occupancy equals $X_{LD}$: $(T_{MD} - T_F) = X_{LD} (T_B-T_F) \approx3.8$ K (Table \ref{tab:2}). Below $T_{MD}$ the $\mathcal{R}T_B$ occupancy vanishes, the $\mathcal{R}T_H$ occupancy increases with $\Theta_{SC}$ (Table \ref{tab:2}), the degeneracy decreases, and the internal energy increases. $\mathcal{R}T_H$ is the lowest state of the liquid, and crystallization at $T_H$ is an adiabatic process in which the latent heat is removed from the liquid. 

This description explains several properties. 

The liquid and the ice are isomorphic. \cite{Soper,NL} The greater degeneracy of the liquid explains the enhanced diffuse scattering. The dipolar energy gain explains the density of the LDL. 

For $T_B \geq T \geq T_{MD}$, equipartition gives $C_{HDL} = 9 \mathcal{R}$ (Table \ref{tab:1}). The kinetic energy $\frac{9}{2} \mathcal{R} (T-T_F)$ explains the NCS data. The occupancy of the degenerate state at $\mathcal{R} T_B$ is proportional to $(T-T_F)^2$, so the O$\cdots$O distances increase and the density decreases by heating, both quadratically. 

$T_{MD} \approx 277$ K corresponds to the maximum density. 

$T_H \approx 226$ K is the lower limit for the homogeneous crystallization, which is observed in the range of $226-232$ K, depending on the experimental conditions. \cite{Ice,AOS} 

For $T_{MD} \geq T \geq T_H$, $C_{SC} = \mathcal{R} (9+ \Theta_{SC}^2 \ln \Omega_{HD})$ (Table \ref{tab:2}) explains the quadratic increase of the heat capacity, from $\approx75$ J.mol.$^{-1}T^{-1}$ at $T_{MD}$ to $\approx103$ J.mol.$^{-1}T^{-1}$ at $T_H$, \cite{AOS} and the quadratic decrease of the density. 

\subsection{\label{sec:2.4} Gaseous water}

The heat of ebullition at $T_B$ and the heat of vaporization below $T_B$ can be treated on the same foot. The isotropic volume expansion of $\approx1700$ of the gas at normal pressure relative to the liquid results from the dissociation of 6 of the 7 dipolar pairs per honeycomb unit. This results in a gas of free dimers \textbf{III}, whose tunneling splitting is $\hbar \omega_{t}/k_B \approx0.94$ K. \cite{OHP} The heat transfer by photons $\omega_{t}$ is temperature independent and the heat capacity is $\frac{9}{2}\mathcal{R}$ as long as the populations of the dissociative states are marginal.  

The energy cost of the liquid-gas transition is 
\begin{equation}\begin{array}{rcl}\label{eq:5} 
\Delta H_{G} (T) & = & 6 D_0 - \mathcal{R} T \displaystyle {\left( 9 + \ln \Omega_{HD} \right )}.\\
\end{array}\end{equation}
The heat of ebullition $\Delta H_G (T_B) = (40575 \pm 700) \ \mathrm{J.mol}^{-1}$ agrees with the empirical value of $(40660 \pm 80) \ \mathrm{J.mol}^{-1}$. \cite{Marsh} Furthermore, the empirical heat of sublimation of ice at $T_F$, namely $\approx51059$ J.mol$^{-1}$, \cite{MK} agrees with $\Delta H_{F} + \Delta H_{G} (T_F) \approx51058$ J.mol$^{-1}$, which means that sublimation is a two-step process through the liquid state. This consolidates the values of $D_0$ and $\Omega_{HD}$.

\subsection{\label{sec:2.5} Water droplets}

Condensation of the vapor over the locally heated water surface can produce long-lived, self-organizing, honeycomb-patterned aggregates composed of equidistant monodisperse droplets as a single layer floating above the surface, and the spatial order is maintained even as the layer moves horizontally. \cite{FFS,FDR,UON,ZKA,AK} The droplet diameter and spacing are $\approx10 \mu$m. The self-assembly mechanism is thought to be a balance between hydrodynamic forces. The repulsive force is attributed to the upward viscous flow of the vapor around the droplets, but there is no convincing explanation for the attractive forces. 

However, droplet are bosons. They emerge from the vapor with no center of mass kinetic energy, so coalescence is avoided. They are subject to two vertical forces, one due to gravity proportional to $r^3$ and the other due to the upward vapor flow proportional to $r^2$, where $r$ is the radius. For a critical value, $r_c$, which depends on $T$, the droplets drift downward at a slow, viscosity-limited rate. The de Broglie wavelength in the horizontal plane encompasses a large number of droplets, and the honeycomb layer of monodisperse spheres at rest with respect to each other results from Bose-Einstein condensation. This layer can move horizontally as a whole without deformation. In the reported experiments, the layer blocks upward vapor flow and prevents multilayer formation. In the open air, however, large-scale condensates in 3-D may explain why droplets cluster in long-lived clouds drifting in the troposphere, instead of scattering toward the state of maximum entropy expected for classical droplets in the absence of mutual attraction. 

\section{Conclusion}

Bose-Einstein condensation at the gas-liquid transition explains the stability of the condensed phases of water, as well as their structures, critical temperatures, latent heats, heat capacities, equipartition violations, low and high density liquids, maximum density, and spectroscopic data with only four observables ($\hbar\omega_t$, $\hbar\omega_1$, $\hbar\omega_\mu$, $D_0$). Condensation provides new insights into the quantum properties of H-bonds, and explains the honeycomb 
structure of droplet layers. There are no arbitrary assumptions, no discontinuity of quantum mechanics, and no decoherence. 

Compared to monoatomic systems, the condensed phases of water are non-superfluid condensates under standard conditions. They extend the size, temperature, and density ranges of condensates by orders of magnitude and condensates should also be relevant to various H-bond containing systems. Although the reasoning should be further tested for properties not considered in this work, it is already likely that Bose-Einstein condensates are a more common state of matter than ever thought. 

\end{document}